\documentclass{ragtime} 

\title[Black hole accretion torus \& magnetic fields]{Simulations of black hole accretion torus in various magnetic field configurations}

\author[M.~Kolo\v{s}, A.~Janiuk]  
{
Martin Kolo\v{s}\at{1,a} and Agnieszka Janiuk\at{2,b} \\
\ins{1} Research Centre for Theoretical Physics and Astrophysics, 
		Institute of Physics  \splitins[1] , Silesian University in Opava,
		Bezru{\v c}ovo n{\'a}m.13, CZ-74601 Opava, Czech Republic\\
\ins{2} Center for Theoretical Physics, 
		Polish Academy of Sciences, \splitins[1]
		Al. Lotnikow 32/46, 02-668 Warsaw, 
		Poland\\
\ins{a}\Email{martin.kolos@fpf.slu.cz}
\ins{b}\Email{agnes@cft.edu.pl}
}

\providecommand{\dif}{\mathrm{d}}
\def\beq{\begin{equation}}
\def\eeq{\end{equation}}
\def\bea{\begin{eqnarray}}
\def\eea{\end{eqnarray}}

\def\ugas{\tilde{u}_{\rm g}}

\def\d{\dif}

\begin{document}

\begin{abstract}
Using axisymetric general relativistic magnetohydrodynamics simulations we study evolution of accretion torus around black hole endowed with different initial magnetic field configurations. Due to accretion of material onto black hole, parabolic magnetic field will develop in accretion torus funnel around vertical axis, for any initial magnetic field configuration.
\end{abstract}



\section{Introduction}

There are two long range forces in physics: gravity and electromagnetism (EM) and both of these forces are crucial for proper description of high energetic processes around black holes (BHs). In realistic astrophysical situations the EM field around BH a is not strong enough ($<10^{18}$~Gs) to really contribute to spacetime curvature and rotating BH can be fully described by standard Kerr metric spacetime. Hence the EM field and matter orbiting around central BH can be considered just as test fields in axially symmetric Kerr spacetime background. While the distribution of matter around central BH can be well described by thin Keplerian accretion disk or thick accretion torus, the exact shape of EM field around BH, i.e. BH magnetosphere is more complicated. In the case of rotating neutron star (pulsar) inclined rotating dipole field is used - such magnetosphere is generated by currents floating on the star surface. In the case of BHs one can assume the magnetosphere will be generated by currents floating around BH inside accretion disk or torus. 

Historically, the question of BH magnetosphere has been solved as vacuum solution of Maxwell equations in curved background. The solution of uniform magnetic field in Kerr metric has been found by Wald \citep{Wald:1974:PHYSR4:}, and can serve as zero approximation to some more realistic BH magnetosphere. In elegant Wald uniform solution one can study combined effect of gravitational and Lorentz force acting on charged mass element. Unfortunately such electrovacuum stationary test field BH magnetosphere has limited astrophysical relevance - material orbiting around BH in the form of plasma should be included. Plasma effect on BH force-free magnetosphere has been included in the well-known work of Blandford \& Znajek \citep{Bla-Zna:1977:MNRAS:}, where also the electromagnetic mechanism of BH rotational energy extraction has been introduced.

Several numerical techniques has been also employed, but the exact shape and intensity of BH magnetosphere, is still not yet properly resolved, although strong connection to the accretion processes is evident \citep{Punsly:BHmag,Meier:TheEngineParadigm}. Simple and elegant solution of uniform magnetic field \citep{Wald:1974:PHYSR4:} could be used as first linear approximation to real BH magnetosphere model, but from GRMHD simulations of accretion processes one can expect the BH magnetosphere has more complicated structure and also changes in time \citep{Tch:2015:ASSL:,Jan-etal:2018:arXiv:}. At small scales the turbulent magnetic field inside accretion disk is very important, since it enables the angular momentum transport inside the accretion disk due to the magnetorotational instability (MRI) \citep{Sap-Jan:2019:APJ:}. At large scales one should use some analytic approximation to real turbulent large scale BH magnetosphere outside the accretion disk. The GRMHD simulations of magnetic field around BH \citep{Nak-etal:2018:APJ:,Por-etal:2019:arXiv:,Lan-etal:2019:APJL:} can provide motivation for heuristic analytic solution for BH magnetosphere. Such analytic BH magnetosphere solution smooth out all small scale and fast time discrepancies and can represent real magnetic field around BH on long times and long scales. Inside this analytic BH magnetosphere one can then study fast physical processes like charged particle jet acceleration \citep{Stu-Kol:2016:EPJC:,Kop-Kar:APJ:2018:} which could be used as model for Ultra-High-Energy Cosmic Rays (UHECR) \citep{Tur-Dad:2019:Uni:, Stu-etal:2020:Universe:}.

\section{Numerical simulation of accretion onto BH}

In this technical section we will introduce equations for our model of accretion torus around BH. The equations will be given geometric units ($G=c=1$) and as compared to the standard Gauss cgs system, the factor $1/\sqrt{4\pi}$ is absorbed in the definition of the magnetic field. Greek indices run through $[0,1,2,3]$ while Roman indices span $[1,2,3]$.  

\subsection{Equations of ideal GRMHD in curved spacetime} 

In our simulations for this proceeding, black hole spin has been neglected, and Schwarzschild geometry has been used for description of central compact object. In the standard coordinates and in the geometric units Schwarzschild metrics takes form
\beq
\d s^{2}= -f(r)\,\d t^2 +f(r)\,\d r^2 +r^2 ( \d\theta^2 + \sin^2 \theta
\,\d\phi^2), \qquad f(r) = 1-\frac{2M}{r}, \label{SchwMetric}
\eeq
where $M$ gravitational mass of the central compact object. In the following, we put $M=1$, i.e., we use dimensionless radial coordinate $r$ and dimensionless time coordinate $t$. In the present paper we restrict our attention to the black hole spacetime region located above the outer event horizon at $r_{\rm h}=2$. 


The plasma orbiting around central Schwarzschild BH will be modeled using ideal GRMHD equations, where electric resistivity, self-gravitational, radiative and all non-equilibrium effects are neglected. The continuity, the four-momentum-energy conservation and induction equations within GRMHD framework are:
\bea
 \left( \rho u^{\mu} \right)_{;\mu} = 0, \qquad
 \left( {T^{\mu}}_{\nu} \right)_{;\mu} = 0, \qquad
 \left( u^\nu b^\mu - u^\nu b^\nu \right)_{;\mu} = 0, \label{idealMHD}
\eea
The stress-energy tensor $T^{\mu\nu}$ is composed of gas and electromagnetic part
\bea
&& T^{\mu\nu}_{\rm gas} = ( \rho + \ugas + p)u^{\mu} u^{\nu}+p g^{\mu\nu}, \,\,\quad\,\,
T^{\mu\nu}_{\rm EM} = b^{2} u^{\mu} u^{\nu}+\frac{1}{2} b^{2} g^{\mu\nu} - b^{\mu} b^{\nu}, \\
&& T^{\mu\nu} = T^{\mu\nu}_{gas}+T^{\mu\nu}_{EM} = ( \rho + \ugas + p + b^2)u^{\mu} u^{\nu}+ \left( p + b^2/2  \right) g^{\mu\nu} - b^{\mu} b^{\nu}.
\label{tensorT}
\eea
Variables in in Eqs. (\ref{idealMHD}-\ref{tensorT}) are: $u^{\mu}$ is gas four-velocity, $\ugas$ is internal gas energy density, $\rho$ is gas density and $p$ denotes gas pressure, and $b^{\mu}$ is the magnetic four-vector. Magnetic four-vector $b^{\mu}$ is related to magnetic field three-vector $B^i$
\beq
  b^t = B^i u^\mu g_{i\mu}, \qquad b^i = (B^i + b^t u^i)/u^t.
\eeq
Strength of the magnetic field in the fluid-frame is given by $B^2= b^\alpha b_\alpha $, we can also define magnetization $\sigma=B^2/\rho$ and the plasma-$\beta$ parameter $\beta=2p/B^2$.

The equation of state (EOS) will be used in the form of ideal gas
$ p = (\hat{\gamma} - 1) \ugas $, where $\hat{\gamma}$ is the adiabatic index \citep{Gam-etal:2003:APJ:}, for simulations with non-adiabatic EOS see \citep{Jan:2017:APJ:}.

\subsection{Initial distribution of matter and EM field around central BH}


Initial conditions in our simulation we will be toroidal perfect fluid configurations of neutral matter around central BH in the form of Polish donut model \citep{Koz-etal:1977:ASTRA:, Abr-etal:1978:ASTRA:}, while for magnetic field we will test various different configurations. The time evolution and relaxation of accretion torus magnetized matter and magnetic field into more realistic configuration will be studied using GRMHD simulation. 

Due to stationarity and axial symmetry of our problem Eq.~(\ref{SchwMetric}) we will assume $\partial_t X =0$ and $\partial_{\varphi} X=0$, with $X$ being a generic spacetime tensor. In the equations for neutral matter distribution Eq.~(\ref{idealMHD}), the continuity equation is identically satisfied and the fluid dynamics is governed by the Euler equation only
\bea
(p+\varrho) u^\alpha \nabla_\alpha u^\gamma +  h^{\beta\gamma}\nabla_\beta p=0 , 
\label{EulerEQ}
\eea
where $\nabla_\alpha g_{\beta\gamma}=0$, $h_{\alpha \beta}=g_{\alpha \beta}+ u_\alpha u_\beta$ is the projection tensor. 

Barotropic equation of state $p=p(\varrho)$ is assumed, and the matter is in orbital motion only $u^{\theta}=0$ and $u^r=0$. The Euler equation (\ref{EulerEQ}) can be written as an equation for the barotropic pressure $p(\varrho)$ as follows \citep{Fis-Mon:1976:APJ:}
\beq
\frac{\partial_{\mu}p}{\varrho+p}=-{\partial_{\mu }W}+\frac{\Omega \partial_{\mu} l}{1-\Omega \ell},  \qquad
W\equiv -\ln\left( -g_{tt} -g_{\phi\phi} \Omega^2 \right) + l_* \Omega, \label{diskEQ}
\eeq
where $l_* = l/1-\Omega l$ is constant through the accretion torus, $\Omega=u^{\phi}/u^{t}$ is the fluid relativistic angular frequency related to distant observers, while $W(r;\ell)$ is the Paczy{\'n}ski-Wiita potential. The fluid equilibrium is regulated by the balance of the gravitational and pressure terms versus centrifugal factors arising due to the fluid rotation and gravitational effects of the BH background.


Relativistic formulation of Maxwell's equations in curved spacetime is
\beq
 \partial_\alpha F_{\mu\nu} + \partial_\nu F_{\alpha\mu} + \partial_\mu F_{\nu\alpha} = 0, \qquad
 \partial_\alpha F^{\alpha\beta} = \mu_0 J^{\beta}. \label{maxwell}
\eeq
where $J^{\beta}$ is electric current four-vector and electromagnetic tensor $F_{\mu\nu}$ is given by
\beq
 F_{\mu\nu} = \partial_\mu A_{\nu} - \partial_\nu A_{\mu},
\eeq
where $A^{\mu}$ is electromagnetic four-vector. Assuming axial symmetry and absence of electric field, the only non-zero component of $A^{\mu}$ will be $A^\phi$, and we can write $ A^\mu = (0,0,0,A^\phi)$.
The first of Maxwell's equations (\ref{maxwell}) is satisfied identically, while the second is giving the equation
\beq
  r^2 \frac{\partial}{\partial r} \left[ \left(1-\frac{2}{r} \right) \frac{\partial}{\partial r} A_{\phi} \right] + \sin\theta \frac{\partial}{\partial \theta} \left(\frac{1}{\sin\theta} \frac{\partial}{\partial \theta} A_{\phi} \right) = - \mu_0 J^\phi \, r^4 \sin^2\theta. \label{rovnice}
\eeq
This equation is Ampere's law, but it can be also wield as special case of Grad---Shafranov equation well known in MHD \citep{Meier:TheEngineParadigm}. Magnetic field three-vector ${\bf B}=(B^{\widehat{r}},B^{\widehat{\theta}},B^{\widehat{\phi}} )$ can be related to four-vector component $A_\phi$ using
\beq \label{magfielddef}
 B^{\widehat{r}} = \frac{1}{\sqrt{-g}} \, A_{\phi,\theta} \quad  
 B^{\widehat{\theta}} = -\left( 1- \frac{2}{r} \right)^{1/2} \frac{1}{r
\sin(\theta)} \, A_{\phi,r}, \quad
 B^{\widehat{\phi}} = 0.
\eeq
Magnetic field ${\bf B}$ is fully specified by electromagnetic four-potential $A^\mu$, see Eq.~(\ref{magfielddef}). While the GRMHD HARM code is using magnetic field ${\bf B}$ in the simulations, it is sometimes more elegant to work with electromagnetic potential $A^\mu$ instead, for example visualization of magnetic field ${\bf B}$ can be easily plotted using contour lines of electromagnetic potential
\beq
 A^\phi (r,\theta) = {\rm const.}
\eeq

\subsection{HARM numerical code} 


HARM (High Accuracy Relativistic Magneto-hydrodynamics) is a conservative shock capturing scheme, for evolving the equations of GRMHD, developed by C.~Gammie et al. \citep{Gam-etal:2003:APJ:} and later improved and parallelized and released as HARM COOL code by A.~Janiuk and her team at CFT PAS in Warsaw \citep{Jan-etal:2018:arXiv:,Pal-Jan-Suk:2019:MNRAS:,Sap-Jan:2019:APJ:}.

The integrated equations for ideal MHD, Eq. (\ref{idealMHD}) can be expressed in general form and then discretizated for each cell from fixed numerical grid
\beq
 \frac{\partial U}{\partial t} + \frac{\partial F^x}{\partial x} = S \qquad
 \frac{\partial <U_i>}{\partial t} 
 + \frac{\partial F^x_{i+1/2}-F^x_{i-1/2}}{\triangle x} = S,
\eeq
where $U$ is a vector of ``conserved variables'', such as particle number density, or energy or momentum, $F^i$ are the fluxes in finite control volume, and $S$ is a vector of source terms. $U$ is conserved in the sense that, if $S=0$, it depends only on fluxes at the boundaries. The vector $P$ is composed of ``primitive'' variables, such as rest-mass density, internal energy density, velocity components, and magnetic field components, which are interpolated to model the flow within zones. $U$ and $F^i$ depend on $P$. Conservative numerical schemes advance $U$, then, depending on the order of the scheme, calculate $P (U)$ once or twice per time step.
Because stationary and spherically symmetric Schwarzschild BH metric (\ref{SchwMetric}) is not changing during whole GRMHD simulation, we can divide the whole space into fixed numerical grid. Moreover for our problem we restrict ourselves to two dimensional ($r,\theta$) subspace. 
In this proceeding we use simulation domain $r\in [0.98\,r_{\rm h},100]$, $\theta\in[0,\pi]$ with resolutions $128\times128$ cells in nonlinear fixed grid.

\section{Results of GRMHD simulations}  



In this short contribution we will try to examine magnetic field structure around accreting BH using GRMHD simulations in HARM COOL code. As initial conditions for our simulations we will use thick accretion torus in hydrodynamic equilibrium which will be immersed into different test magnetic field configurations. 

Standard settings setup used in GRMHD simulations are: thick accretion torus around central rotating Kerr BH with dimensionless spin parameter $a = 0.9375$; torus inner radius at $r_{\rm in} = 6$ and the torus density maximum at $r_{\rm max}=12$ \citep{Gam-etal:2003:APJ:,Por-etal:2019:arXiv:}. Angular momentum distribution inside the torus is prescribed by Eq.~(\ref{diskEQ}) \citep{Fis-Mon:1976:APJ:}. In this proceeding we would like to simulate accretion torus around nonrotating Schwarzschild BH, but with similar central density as in Kerr BH case, hence for our simulation we use accretion torus with inner radius at $r_{\rm in} = 8$ and density maximum at $r_{\rm max}=16$.

As seed for torus inhomogeneities, we perturb thermal pressure inside torus by $p^* = p\,(1+X_p)$ function, where $X_p$ is a uniformly distributed random variable between $-0.02$ and $0.02$. We use ideal gas equation of state with an adiabatic index of $\hat{\gamma}=4/3$. 
We will run the simulations till final time $t=10^4$, which is around 30 orbits around black hole for matter form accretion torus. Since the accretion torus is in differential rotation we will relate torus orbital period to torus density maximum - one free test particle circular orbit around BH at $r=16$ take $t\sim363$ time in geometric units used in our simulation. 

\begin{figure}
\includegraphics[width=0.329\textwidth]{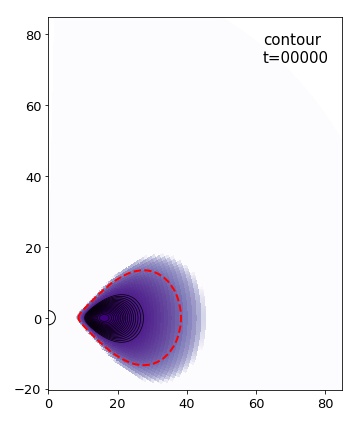}
\includegraphics[width=0.329\textwidth]{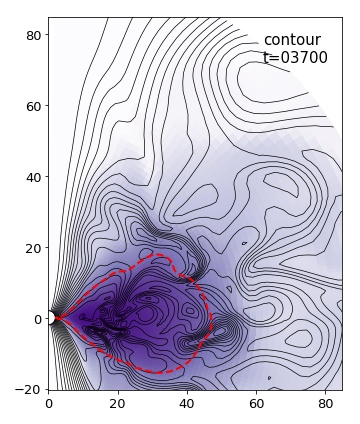}
\includegraphics[width=0.329\textwidth]{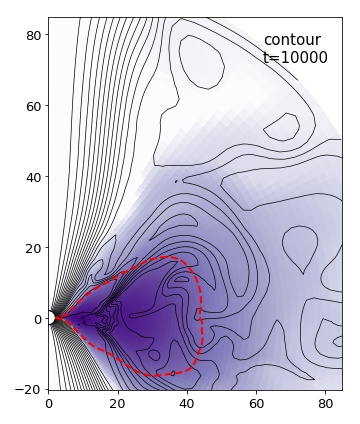}
\caption{
Stages of numerical GRMHD simulation of accretion torus in  magnetic field following by matter density contours. Initial, middle and final stage of numerical GRMHD simulation of accretion torus in uniform magnetic field.
Only 2D sections of full axially symmetric accretion torus are plotted, with $x$ on horizontal axis and $z$ (axis of BH rotation) on vertical axis. Black curves represent magnetic field lines, black circle at the origin of coordinates represent BH horizon.
Different shades of blue color represents logarithmic density of matter form accretion torus - the region with 99\% of maximal density (accretion torus itself) is bounded by thick red dashed curve, while the region with $10^{-6} \rho_{\rm max}$ (wind around accretion torus) is bounded by thick dashed curve. Time of the simulations in the unis of $M$ is given in the right up corner.
\label{Fcon}
}
\end{figure} 

\begin{figure}
\includegraphics[width=0.329\textwidth]{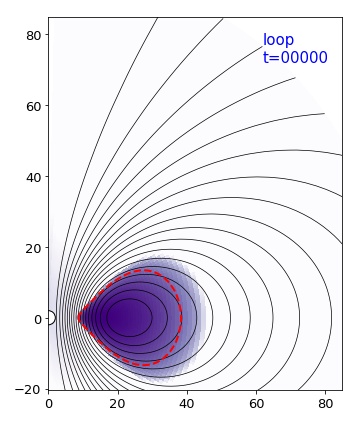}
\includegraphics[width=0.329\textwidth]{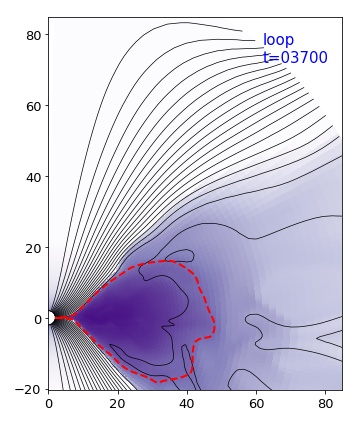}
\includegraphics[width=0.329\textwidth]{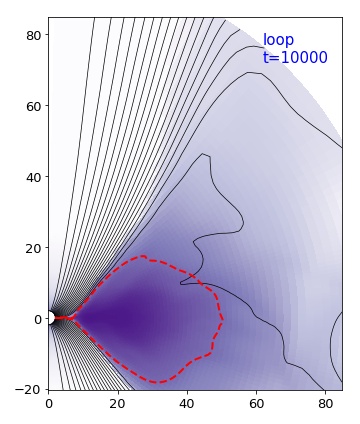}
\caption{Stages of numerical GRMHD simulation of accretion torus in  magnetic field given by current loop. \label{Floop}}
%
\includegraphics[width=0.329\textwidth]{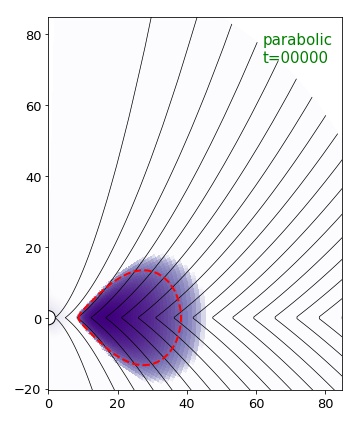}
\includegraphics[width=0.329\textwidth]{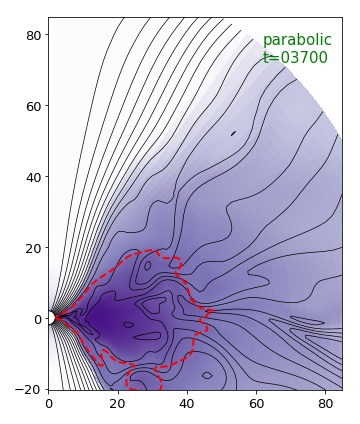}
\includegraphics[width=0.329\textwidth]{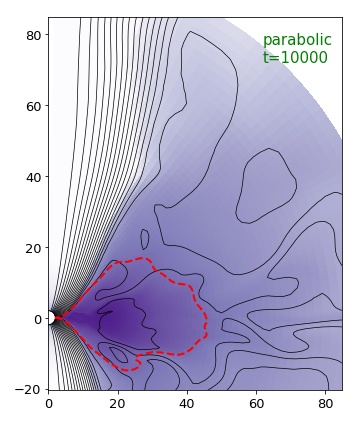}
\caption{Stages of GRMHD simulation of accretion torus in parabolic mag. field. \label{Fpar}}
\includegraphics[width=0.329\textwidth]{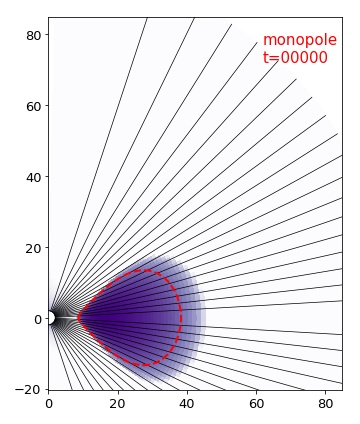}
\includegraphics[width=0.329\textwidth]{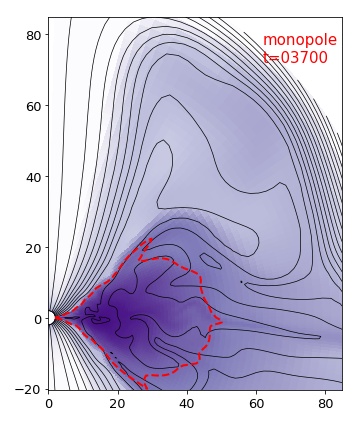}
\includegraphics[width=0.329\textwidth]{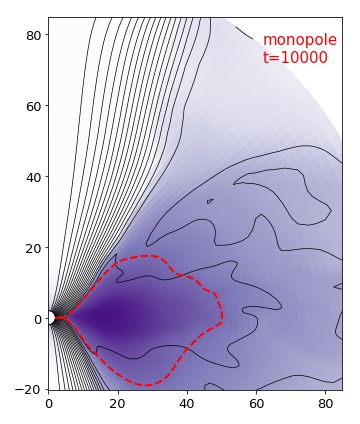}
\caption{Stages of GRMHD simulation of accretion torus in split monopole mag. field. \label{Fmono}}
\end{figure} 

\begin{figure}
\includegraphics[width=0.329\textwidth]{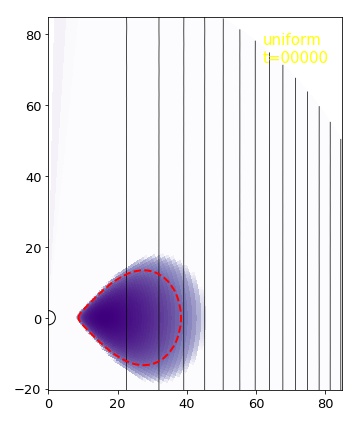}
\includegraphics[width=0.329\textwidth]{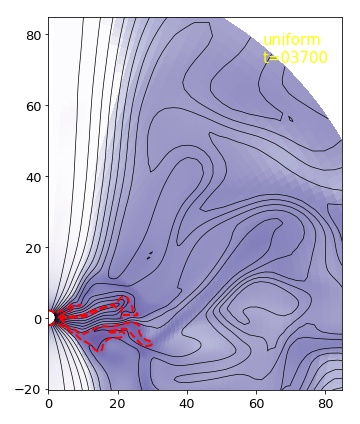}
\includegraphics[width=0.329\textwidth]{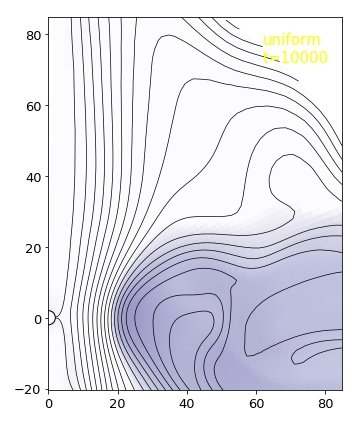}
\caption{Stages of numerical GRMHD simulation of accretion torus in uniform magnetic field.
Some problematic behavior is observed for uniform magnetic field configuration. Contrary to the previous four cases, the uniform magnetic field will disturb the accretion torus so much, that he will be quickly swallowed by BH. After ten orbits (middle subfigure) only strongly destroyed torus structure can be visible in the model. 
\label{Funi}
}
\end{figure}

Different magnetic field configurations will be tested as initial EM field in which the accretion torus will be immersed. Some of them are solution of vacuum Maxwell equation in curved spacetime Eq.~(\ref{rovnice}), some of them are just heuristic approximation. We will start with standard initial setting for HARM torus simulation with poloidal magnetic field following the {\bf contours} of matter \citep{Gam-etal:2003:APJ:,Por-etal:2019:arXiv:}. Here the magnetic field lines are closed curves focused around center at maximal torus density radius, see Fig.~\ref{Fcon}. Magnetic field strength is set to be $\beta = 2 p_{\rm max}/(B^2)_{\rm max}=100$, where $p_{\rm max}$ and $B_{\rm max}$ are pressure and magnetic field magnitude at torus density center ($r=16, \theta=\pi/2$).

Another magnetic field configuration with closed magnetic field lines is magnetic field generated by current {\bf loop} located in equatorial plane at given radii $r=R$ \citep{Pet:1974:PRD:}, see Fig.~\ref{Floop}. We will use simplified formula (1st leading term in expansion) for this Petterson current loop magnetic field \citep{Kol:2017:RAGtime:}
\beq
A_{\phi} = B \, \sqrt{32} \, R^3  \frac{r \sin\theta}{\left(R^2+r^2\right)^{3/2}},
\label{Aloop}
\eeq
where we set the magnetic parameter to be $B=0.003$. Term $\sqrt{32}\, R^3$ is normalization factor which has been added to normalize magnetic field magnitude at disk pressure maximum ($r=16$) to value $B$. All following analytical magnetic field configurations will be normalized in this way and use magnetic field parameter $B=0.003$.

GRMHD simulations of accretion processes around central BH \citep{Nak-etal:2018:APJ:,Por-etal:2019:arXiv:} are giving {\bf parabolic} magnetic field as filed configuration inside the accretion torus funnel. Analytic formula for parabolic magnetic field is given by
\beq
A_{\phi} = B \, \frac{128}{5} \, r^k ( 1 - | \cos(\theta) | ), \label{Apar}
\eeq
where we use coefficient $k=0.75$ and add normalization factor $128/5$. Parabolic magnetic field with its open field-lines is plotted in Fig.~\ref{Fpar}.

Well know split {\bf monopole} magnetic field, already studied using test particle dynamic approach in \citep{Bla-Zna:1977:MNRAS:, Kol-Bar-Jur:2019:RAGtime:}, is given by
\beq
A_{\phi} = B \cdot 256 \, ( 1 - | \cos(\theta) | ), \label{Amono}
\eeq
where the magnetic field lines are straight radial lines pointing from the BH above equatorial plane, while to the BH below eq. plane, see Fig.~\ref{Fmono}. This magnetic configuration is solution of Maxwell equations (\ref{rovnice}), but current sheet in equatorial plane is needed.

Classical Wald {\bf uniform} magnetic solution \citep{Wald:1974:PHYSR4:} is given by
\beq
A_{\phi} = B \, r^2 \sin(\theta)^2, \label{Auni}
\eeq
where the magnetic field lines are straight lines parallel with $z$-axis, see Fig.~\ref{Funi}. This magnetic field configuration is solution of vacuum Maxwell equation (\ref{rovnice}). Uniform, monopole, parabolic and loop magnetic field Eqs. (\ref{Aloop}-\ref{Auni}) has been normalized - the Lorentz force acting on charged particle will be the same in all different magnetic configurations at point $r=16, \theta=\pi/2$ (accretion torus pressure maximum). 

\begin{figure}
\includegraphics[width=\textwidth]{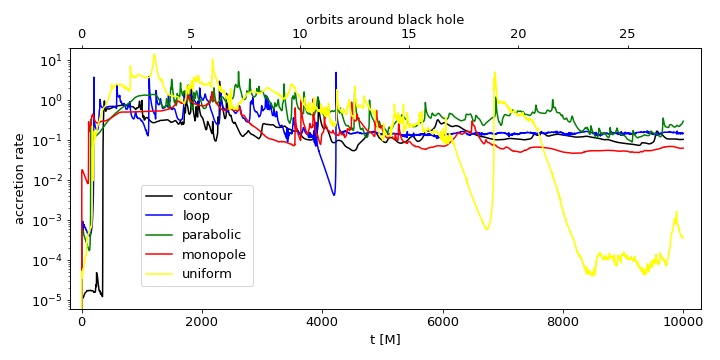}
\caption{Mass accretion rate onto magnetized BH with different initial magnetic field configurations - see timeframes plotted at Figs. \ref{Fcon}-\ref{Funi}. \label{Facc}}
\end{figure}

As it could be seen from simulation results, presented in Figs. \ref{Fcon}-\ref{Funi}, from the beginning of the simulation till circa fifteen orbital periods ($t\sim5000$) the accretion torus experience turbulent regime, when our tested magnetic field configurations are trying to reach some relaxed state. Contrary to the heuristic initial magnetic field configuration, the final relaxed state will be solution of full set of ideal MHD equations (\ref{idealMHD}),  and hence can represent proper realistic BH magnetosphere model. From twenty orbits ($t\sim7500$) till the end of the simulation ($t=10^4$) the accretion flow onto BH is stable and the accretion torus with magnetic field is not changing dramatically, see Fig.~\ref{Facc}. 
 
The axially symmetric GRMHD simulations for our fife different magnetic field configurations shows similar time evolution. After fifteen orbits they will all evolved into the more or less similar state with chaotic and turbulent magnetic field inside accretion torus and regular parabolic magnetic field in accretion torus funnel. Only for uniform magnetic field initial configuration we can see different evolution. In this case the accretion torus structure is quickly destroyed and only some low density corona will remain around central BH. 

In all tested cases the initial magnetic field configuration is quickly erased. After some time, one can distinguish in relaxed state solution new formed  regions which can be classified according to the magnetic field shape and matter distribution. 
\vspace{-0.5cm}
\begin{description}
\item[Torus] - Where the matter density in high $\rho\sim\rho_{\rm c}$ and  matter dominate over the magnetic field. Magnetic field inside torus is turbulent and chaotic and contributes to the accretion disk viscosity through magnetorotational instability.
\item[Corona] - Where the matter density is much lower $\rho\leq\rho_{\rm c}$, but matter still dominate over the still turbulent magnetic field.
\item[Jet funnel] - Where the matter component is missing $\rho\leq10^{-6}\rho_{\rm c}$ and regular magnetic field with parabolic shape dominate the region. 
\end{description}
Matter form corona low density region can be easily ionized at the jet funnel/corona boundary and description of collisionsless charged test particle dynamic in given magnetic field can be well applied in this jet funnel region. Funnel region with parabolic magnetic field will be important for charged particles acceleration to ultra-relativistic velocities and production of UHECR \citep{Stu-Kol:2016:EPJC:,Kop-Kar:APJ:2018:,Tur-Dad:2019:Uni:,Stu-etal:2020:Universe:}.

\section{Conclusions}

In this short text we examined five different magnetic field configuration and tested their evolution during matter accretion process. Simple asymmetric torus orbiting around central BH has been penetrated by made up magnetic field configurations and using GRMHD numerical simulations we studied matter accretion onto BH and tested magnetic field evolution. Due to accretion of material onto BH, regular magnetic field with parabolic shape has develop in accretion torus funnel around vertical axis. 
Turbulent and chaotic magnetic field inside torus will redistribute angular momentum inside torus, create corona around the torus and will initiate BH accretion process.

In future work we would like to use GRMHD numerical simulations not only to calculate exact shape of BH magnetosphere but also to provide the distribution of different types of elementary particles and their velocities inside accretion torus \citep{Jan-etal:2018:arXiv:}. At the corona/jet funnel boundary, charged particles from the quasi-neutral accretion torus will no longer feel the pressure forces, and they can start to move under the combined influence of gravity and EM Lorentz force only. Hence charged particles can be accelerated and they can escape with ultrarelativistic velocities along magnetic filed lines toward infinity \citep{Stu-etal:2020:Universe:}. Knowing the charged particles radiation losses over their full path to Earth atmosphere \citep{Tur-etal:APJ:2018:}, one could be able to calculate the distribution of UHECR particles in the shower hitting Earth surface.


\def\prc{Phys. Rev. C}
\def\pre{Phys. Rev. E}
\def\prd{Phys. Rev. D}
\def\jcap{Journal of Cosmology and Astroparticle Physics}
\def\apss{Astrophysics and Space Science}
\def\mnras{Monthly Notices of the Royal Astronomical Society}
\def\apj{The Astrophysical Journal}
\def\aap{Astronomy and Astrophysics}
\def\actaa{Acta Astronomica}
\def\pasj{Publications of the Astronomical Society of Japan}
\def\apjl{Astrophysical Journal Letters}
\def\pasa{Publications Astronomical Society of Australia}
\def\nat{Nature}
\def\physrep{Physics Reports}
\def\araa{Annual Review of Astronomy and Astrophysics}
\def\apjs{The Astrophysical Journal Supplement}
\def\aapr{The Astronomy and Astrophysics Review}
\def\procspie{Proceedings of the SPIE}

\def\mdash{---}


\end{document}